\documentclass{article}
\usepackage{spconf,amsmath,graphicx}

\usepackage{booktabs}
\usepackage{amssymb,amsmath,bm}

\usepackage[usenames,dvipsnames]{xcolor}
\usepackage[bookmarks, colorlinks, breaklinks]{hyperref}

\definecolor{darkred}{rgb}{0.5,0,0}
\definecolor{lightblue}{rgb}{0,0.4,0.8}
\definecolor{darkgreen}{rgb}{0,0.5,0}
\hypersetup{colorlinks=true, linkcolor=blue, urlcolor=blue, citecolor=darkgreen}

\DeclareMathOperator*{\argmax}{arg\,max}

\usepackage{siunitx}
\newcommand{\mybff}[1]{\bm{#1}}

\title{FASTPITCH: PARALLEL TEXT-TO-SPEECH WITH PITCH PREDICTION}
\name{Adrian \L{}a\'ncucki}  %
\address{NVIDIA Corporation}
\begin{document}
\maketitle
\begin{abstract}
We present FastPitch, a fully-parallel text-to-speech model based on FastSpeech, conditioned on fundamental frequency contours.
The model predicts pitch contours during inference.
By altering these predictions, the generated speech can be more expressive, better match the semantic of the utterance, and in the end more engaging to the listener.
Uniformly increasing or decreasing pitch with FastPitch generates speech that resembles the voluntary modulation of voice.
Conditioning on frequency contours improves the overall quality of synthesized speech, making it comparable to state-of-the-art.
It does not introduce an overhead, and FastPitch retains the favorable, fully-parallel Transformer architecture, with over $900\times$ real-time factor for mel-spectrogram synthesis of a typical utterance.
\end{abstract}
\section{INTRODUCTION}
Recent advances in neural text-to-speech (TTS) enabled real-time synthesis of naturally sounding, human-like speech.
Parallel models are able to synthesize mel-spectrograms orders of magnitude faster than autoregressive ones,
either by relying on external alignments~\cite{ren2019}, or aligning themselves~\cite{kim2020}.
TTS models can be conditioned on qualities of speech
such as linguistic features and fundamental frequency~\cite{oord2016}.  %
The latter has been repeatedly shown to improve the quality
of neural, but also concatenative models~\cite{fernandez2015,kons2018}.  %
Conditioning on $F_0$ %
is a common approach to adding singing capabilities~\cite{valle2019mellotron}.  %
or adapting to other speakers~\cite{kons2018}.

In this paper we propose FastPitch, a feed-forward model based on FastSpeech that improves the quality of synthesized speech.
By conditioning on fundamental frequency estimated for every input symbol,
which we refer to simply as a pitch contour,
it matches the state-of-the-art autoregressive TTS models.
We show that explicit modeling of such pitch contours addresses
the quality shortcomings of the plain feed-forward Transformer architecture.
These most likely arise from collapsing different pronunciations
of the same phonetic units in the absence of enough linguistic information in the textual input alone.
Conditioning on fundamental frequency also improves convergence,
and eliminates the need for knowledge distillation of mel-spectrogram targets used in FastSpeech.
We would like to note that a concurrently developed FastSpeech 2~\cite{ren2020}
describes a similar approach.

Combined with WaveGlow~\cite{prenger2019}, FastPitch is able to synthesize mel-spectrograms 
over 60$\times$~faster than real-time,
without resorting to kernel-level optimizations~\cite{ping2018}.
Because the model learns to predict and use pitch in a low resolution of one value for every input symbol,
it makes it easy to adjust pitch interactively, enabling practical applications
in pitch editing.
Constant offsetting of $F_0$ with FastPitch
produces naturally sounding low- and high-pitched variations of voice
that preserve the perceived speaker identity.
We conclude that the model learns to mimic the action of vocal chords,
which happens during the voluntary modulation of voice.

\section{MODEL DESCRIPTION}
\begin{figure}[!htb]
    \centering
    \includegraphics[width=0.8\linewidth]{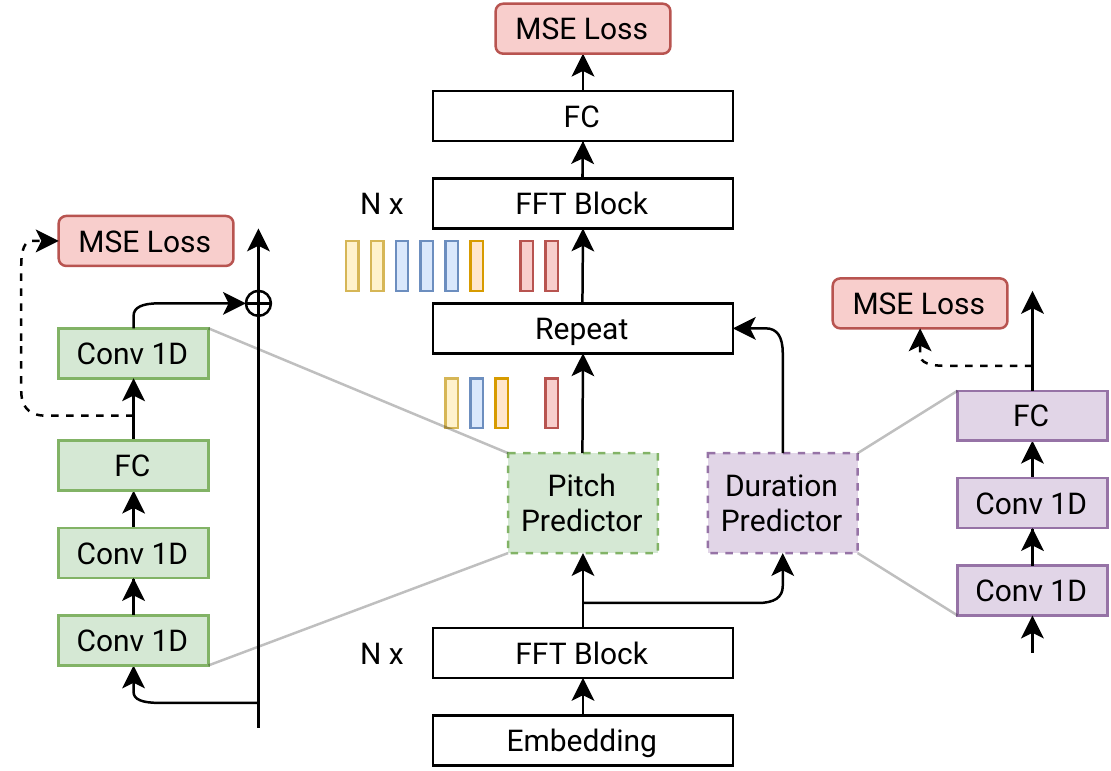}
    \caption{\textbf{Architecture of FastPitch} follows FastSpeech~\cite{ren2019}.
             A single pitch value is predicted for every temporal location.}
    \label{fig:model}
\end{figure}

The architecture of FastPitch is shown in Figure~\ref{fig:model}. It is based on FastSpeech
and composed mainly of two feed-forward Transformer (FFTr) stacks~\cite{ren2019}.
The first one operates in the resolution of input tokens, the second one in the resolution of the output frames.
Let $\mybff{x}=(x_1,\ldots,x_n)$ be the sequence of input lexical units, and $\mybff{y}=(y_1,\ldots,y_t)$
be the sequence of target mel-scale spectrogram frames.
The first FFTr stack produces the hidden representation
$\mybff{h} = \text{FFTr}(\mybff{x})$.
The hidden representation $\mybff{h}$ is used to make predictions about the duration and average pitch of every character with a 1-D CNN

\begin{equation}
  \hat{\mybff{d}} = \text{DurationPredictor}(\mybff{h}),\quad
  \hat{\mybff{p}} = \text{PitchPredictor}(\mybff{h}),
\end{equation}
where $\hat{\mybff{d}}\in\mathbb{N}^n$ and $\hat{\mybff{p}}\in\mathbb{R}^n$.
Next, the pitch is projected to match the dimensionality of the hidden representation
$\mybff{h}\in\mathbb{R}^{n\times d}$
and added to $\mybff{h}$.
The resulting sum $\mybff{g}$ is discretely up-sampled and passed
to the output FFTr, which produces the output mel-spectrogram sequence
\begin{align}
\begin{split}
  \mybff{g} & = \mybff{h} + \text{PitchEmbedding}(\mybff{p})\\
  \hat{\mybff{y}} & = \text{FFTr}(
  [
    \underbrace{g_1, \ldots, g_1}_{d_1},
    \ldots
    \underbrace{g_n, \ldots, g_n}_{d_n}
  ]).
\end{split}
\end{align}

Ground truth $\mybff{p}$ and $\mybff{d}$ are used during training,
and predicted $\hat{\mybff{p}}$ and $\hat{\mybff{d}}$ are used during inference.
The model optimizes mean-squared error (MSE) between the predicted and ground-truth modalities
\begin{equation}
  \mathcal{L} = \lVert \hat{\mybff{y}} - \mybff{y}\rVert^2_2 + 
    \alpha      \lVert \hat{\mybff{p}} - \mybff{p}\rVert^2_2 + 
    \gamma      \lVert \hat{\mybff{d}} - \mybff{d}\rVert^2_2.
\end{equation}

\subsection{Duration of Input Symbols}
Durations of input symbols are estimated with a Tacotron 2 model trained on LJSpeech-1.1~\cite{ljspeech17}.
Let $\mybff{A}\in\mathbb{R}^{n\times t}$ be the final Tacotron 2 attention matrix.
The duration of the $i$th input symbol is~\cite{ren2019}
$d_i = \sum_{c=1}^t [\argmax_r \mybff{A}_{r,c} = i ]$.
Because Tacotron 2 has a single attention matrix, we do not need to choose between attention heads,
as it would be necessary with a multi-head Transformer model.

\begin{figure}[!h]
    \centering
    \includegraphics[width=1.0\linewidth]{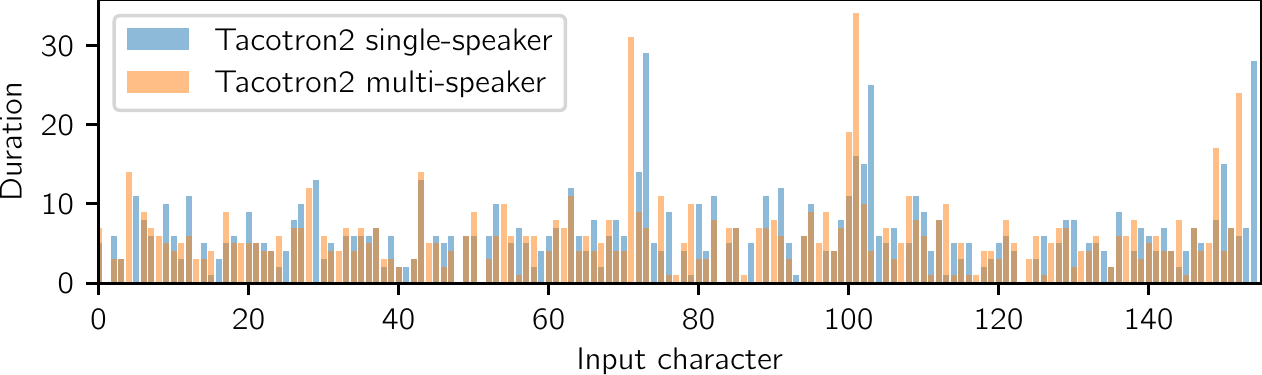}
    \caption{\textbf{Varying character durations} extracted with different Tacotron 2 models
             allow to train FastPitch of similar quality (cf. FastPitch MOS scores in Table~\ref{tab:mos} and Table~\ref{tab:mos-multi}).}
    \label{fig:misaligned}
\end{figure}

FastPitch is robust to the quality of alignments. We observe that durations extracted with distinct
Tacotron 2 models tend to differ (Figure~\ref{fig:misaligned}),
where the longest durations have approximately the same locations, but may be assigned to different characters.
Surprisingly, those different alignment models produce FastPitch
models, which synthesize speech of similar quality.

\subsection{Pitch of Input Symbols}
We obtain ground truth pitch values through acoustic periodicity detection
using the accurate autocorrelation method~\cite{boersma1993}.
The windowed signal is calculated using Hann windows.
The algorithm finds an array of maxima of the normalized
autocorrelation function,
which become the candidate frequencies.
The lowest-cost path through the array of candidates is calculated with the Viterbi algorithm.
The path minimizes the transitions between the candidate frequencies.
We set windows size to match the resolution of training mel-spectrograms, to get one $F_0$ value for every frame.

\begin{figure}
    \centering
    \includegraphics[scale=0.57]{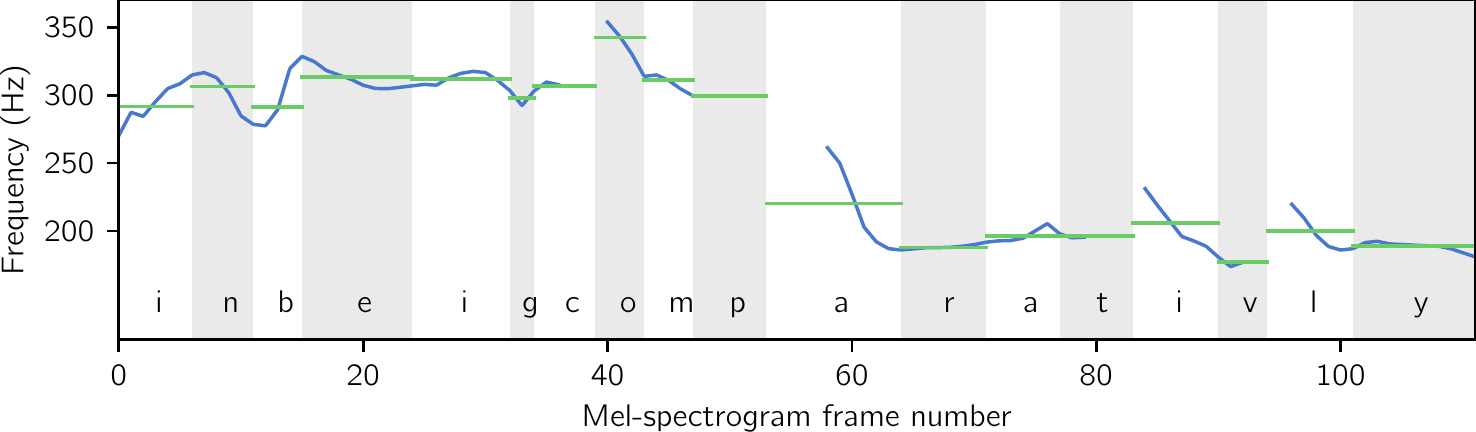}
    \caption{\textbf{Fundamental frequency} estimated for the utterance \textit{In being comparatively}.
    Raw values are shown in blue, values averaged over input characters in green.
    }
    \label{fig:pitch}
\end{figure}

$F_0$ values are averaged over every input symbol using the extracted durations $\mybff{d}$ (Figure~\ref{fig:pitch}).
Unvoiced values are excluded from the calculation.
For training, the values are standardized to mean of 0 and standard deviation of 1.
If there are no voiced $F_0$ estimates for a particular symbol, its pitch is being set to 0.
We have not seen any improvements from modeling $F_0$ in log-domain.

Following~\cite{kons2018}, we tried averaging to three pitch values per every symbol,
in the hope to capture the beginning, middle and ending pitch for every symbol.
However, the model was judged inferior (Section~\ref{sec:eval}).

\section{RELATED WORK}
Developed concurrently to our model, FastSpeech 2~\cite{ren2020}
has a different approach to conditioning on $F_0$ and uses phoneme inputs.
The predicted contour has a resolution of one value for every mel-spectrogram frame, discretized to 256 frequency values.
Additionally, the model is conditioned on energy.
In FastPitch, the predicted contour has one value for every input symbol.
In the experiments this lower resolution made it easier for the model to predict the contour,
and for the user to later modify the pitch interactively.
We found that this resolution is sufficient for the model to discern between different ways of pronouncing a grapheme during training.
In addition, conditioning on higher formants might increase slightly
the quality, capturable by pairwise comparisons (Section~\ref{sec:glicko}).

The predominant paradigm in text-to-speech is two-stage synthesis:
first producing mel-scale spectrograms from text,
and then the actual sound waves with a vocoder model~\cite{ping2018,shen2018,li2019}.
In attempts to speed up the synthesis, parallel models have been
explored. In addition to Transformer models investigated in this
work~\cite{ren2019}, 
convolutional GAN-TTS~\cite{bikowski2019high} is able to synthesize
raw audio waveforms with state-of-the-art quality.
It is conditioned on linguistic and pitch features.

The efforts in parallelizing existing models include duration prediction
similar to FastSpeech, applied to Tacotron~\cite{forwardtacotron20},
WaveRNN~\cite{chengzhu2019durian},
or a flow-based model~\cite{kim2020}.
Explicit modeling of duration 
typically use dynamic programming algorithms associated with
inference and training of HMMs. Glow-TTS aligns with Viterbi paths,
and FastSpeech has been improved with a variant of the forward-backward algorithm~\cite{zeng2020}.

Explicit neural modeling of pitch was
introduced alongside a neural TTS voice conversion model~\cite{kons2018},
which shares similarities with other models from IBM Research~\cite{fernandez2015}.  %
An LSTM-based Variational Autoencoder generation network modeled prosody,
and pitch was calculated with a separate tool prior to the training.
Prosody information was encoded in vectors of four values: log-duration, start
log-pitch, end log-pitch, and log-energy.

\section{EXPERIMENTS}
The source code with pre-trained checkpoints\footnote{\url{https://github.com/NVIDIA/DeepLearningExamples/tree/master/PyTorch/SpeechSynthesis/FastPitch}},
and synthesized samples\footnote{\url{https://fastpitch.github.io/}}
are available on-line.
We synthesize waveforms for evaluation with pre-trained WaveGlow~\cite{prenger2019}.

The model is trained on the publicly available LJSpeech 1.1 dataset~\cite{ljspeech17} which contains
approximately 24 hours of single-speaker speech recorded at \SI{22050}{\hertz}.
We manually correct wrong transcriptions for samples LJ034-0138 and LJ031-0175
discovered during a data inspection with a speech recognition model.
We are cautious to use the same train/dev/test split as our WaveGlow model.
This detail is easy to miss when using a pre-trained model,
and can leak the training data during evaluation and inflate the results.

Parameters of the model mostly follow FastSpeech~\cite{ren2019}.
Each FFTr layer is composed of a 1-D conv
with kernel size 3 and 384/1536 input/output channels, ReLU activation,
a 1-D conv with kernel size 3 and 1536/384 input/output filters,
followed by Dropout and Layer Norm.
Duration Predictor and Pitch Predictor have the same architecture:
a 1-D conv with kernel size 3 and 384/256 channels,
and a 1-D conv with 256/256 channels, each followed by ReLU, Layer Norm and Dropout layers.
The last layer projects every 256-channel vector to a scalar.
Dropout rate is 0.1, also on attention heads.

All described models were trained on graphemes.
Training on phonemes leads to a similar quality of a model, with either
Tacotron 2 or Montreal Forced Aligner~\cite{mcauliffe2017} durations.
However, the mixed approach of training on phonemes and graphemes~\cite{kastner2019}
introduced unpleasant artifacts.

FastPitch has been trained on $8\times$ NVIDIA V100 GPUs with 32 examples per GPU
and automatic mixed precision~\cite{micikevicius2018}.
The training converges after 2~hours, and full training takes 5.5 hours.
We use the LAMB optimizer~\cite{you2020} with learning rate $0.1$, $\beta_1=0.9$, $\beta_2=0.98$, and $\epsilon=\num{1e-9}$.
Learning rate is increased during $1000$ warmup steps, and then decayed according to the Transformer schedule~\cite{vaswani2017}.
We apply weight decay of \num{1e-6}.

\subsection{Evaluation}
\label{sec:eval}
We have compared our FastPitch model with Tacotron 2 (Table~\ref{tab:mos}).
The samples have been scored on Amazon Turk with the Crowdsourced Audio Quality Evaluation Toolkit~\cite{caqe2016}.
We have generated speech from the first 30 samples
from our development subset of the LJSpeech-1.1.
At least 250 scores have been gathered per every model,
with the total of 60 unique Turkers participating in the study.
In order to qualify, the Turkers were asked to pass a hearing test.
\begin{table}[h]
    \centering
    \caption{\textbf{Mean Opinion Scores} with 95\% confidence intervals. Both models were trained on grapheme inputs.}
    \label{tab:mos}
    \begin{tabular}{l l}
      \toprule
        Model                         & MOS\\
      \midrule
        Tacotron 2    (Mel + WaveGlow) & $3.946 \pm 0.134$\\
        FastPitch    (Mel + WaveGlow) & $4.080 \pm 0.133$\\
      \bottomrule
    \end{tabular}
\end{table}

\subsubsection{Pairwise Comparisons}
\label{sec:glicko}
Generative models pose difficulties for hyperparameter tuning.
The quality of generated samples is subjective, and running large-scale studies
time-consuming and costly.
In order to efficiently rank multiple models, and avoid score drift when the developer
scores samples over a long period of time, we have investigated the approach of blindly comparing pairs of samples.
Pairwise comparisons allow to build a global ranking, assuming that skill ratings are transitive~\cite{baumann2017}.

\begin{figure}[!htb]
  \centering
    \includegraphics[scale=0.65]{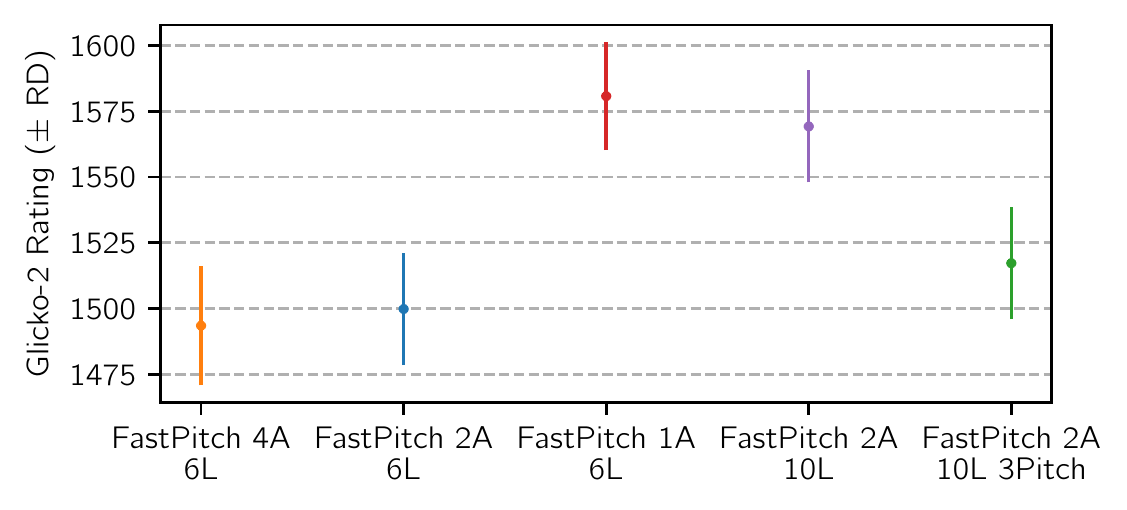}
    \caption{\textbf{Glicko-2 ranking}.
             The models have different numbers of attention heads (A), layers (L), and
             pitch averaged to three values per input symbol instead of one (3Pitch).}
    \label{fig:glicko}
\end{figure}

In an internal study over 50 participants scored randomly selected pairs of samples.
Glicko-2 rating system~\cite{glickman2013},
known from rating human players in chess, sports and on-line games,
but also in the context of automatic scoring of generative models~\cite{olsson2018skill},
was then used to build a ranking based on those scores (Figure~\ref{fig:glicko}).
FastPitch variations with 1, 2 and 4 attention heads, 6 and 10 transformer layers,
and pitch predicted in the resolution of one and three values per input token were compared.
In addition, this rating method has proven useful during development
in tracking multiple hyperparameter settings,
even with a handful of evaluators.

\subsubsection{Multiple Speakers}
FastPitch is easy to extend to multiple speakers. We have trained a model
on the LJSpeech-1.1 dataset with additional internal training data coming from two female speakers:
speaker 1 (8330 samples with the total of $\SI{13.6}{h}$), and speaker 2 ($18995$ samples with the total of $\SI{17.3}{h}$).
We condition the model on the speaker by adding a global speaker embedding
to the input tokens $\mybff{x}$.
To compare, we have chosen multi-speaker Tacotron 2 and Flowtron~\cite{valle2021flowtron}.
The latter is an autoregressive flow-based model. All models have been trained on the same data,
and the multi-speaker Tacotron 2 has been used
to extract training alignments for FastPitch. The results are summarized in Table~\ref{tab:mos-multi}.
\begin{table}[!hb]
    \centering
    \caption{\textbf{Multi-speaker Mean Opinion Scores} with 95\% confidence intervals,
    evaluated on the LJSpeech dev samples}
    \label{tab:mos-multi}
    \begin{tabular}{l l}
      \toprule
        Model                         & MOS\\
      \midrule
        Tacotron 2    (Mel + WaveGlow) & $3.707 \pm 0.218$\\
        Flowtron     (Mel + WaveGlow) & $3.882 \pm 0.159$\\
        FastPitch    (Mel + WaveGlow) & $4.071 \pm 0.164$\\
      \bottomrule
    \end{tabular}
\end{table}

\subsection{Pitch Conditioning and Inference Performance}
A predicted pitch contour can be modified during inference to control certain perceived qualities of the generated speech.
It can be used to increase or decrease $F_0$, raise expressiveness and variance of pitch.
The audio samples accompanying this paper demonstrate
the effects of increasing, decreasing or inverting the frequency
around the mean value for a single utterance, and interpolating between speakers for the multi-speaker model.
We encourage the reader to listen to them.

\begin{figure}[!ht]
      \centering
      \begin{minipage}{0.495\textwidth}
      \centering
      \includegraphics[scale=0.65]{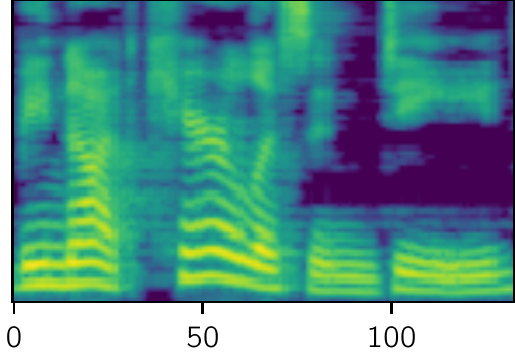}
      \includegraphics[scale=0.65]{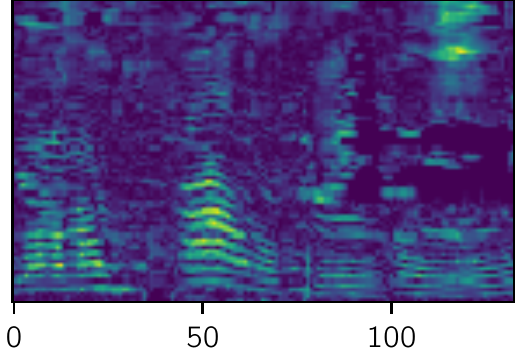}\\
      (a) $F_0$ shifted uniformly by $-\SI{50}{\hertz}$
      \end{minipage}
      \begin{minipage}{0.495\textwidth}
      \centering
      \includegraphics[scale=0.65]{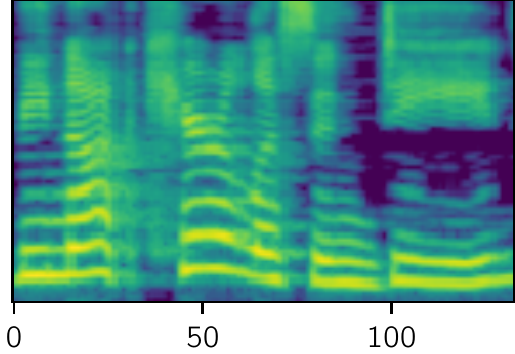}
      \includegraphics[scale=0.65]{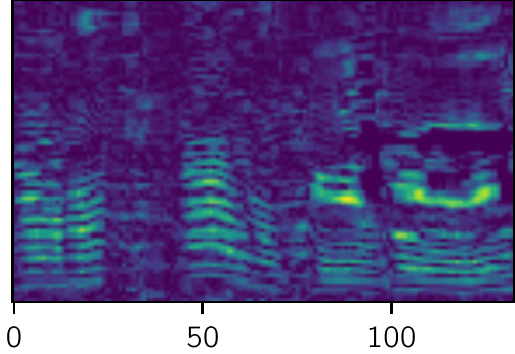}\\
      (b) $F_0$ shifted uniformly by $+\SI{50}{\hertz}$
      \end{minipage}
      \caption{\textbf{Shifting $F_0$ with FastPitch} by adding a constant to the predicted pitch $\hat{\mybff{p}}$ during inference. Pairs are displayed: a  shifted spectrogram, and the absolute difference between shifted and unshifted spectrograms.}
      \label{fig:pitch_shift}
  \end{figure}

Figure~\ref{fig:pitch_shift} shows an example of shifting the frequency by $\SI{50}{Hz}$.
Compared to simple shifting in the frequency domain, FastPitch
preserves the perceived identity of the speaker,
and models the action of vocal chords that happens during voluntary
modulation of voice.

Inference performance measurements were taken on NVIDIA A100 GPU
in FP16 precision and TorchScript-serialized models.
With batch size 1, the average real-time factor (RTF)
for the first 2048 utterances from LJSpeech-1.1 training set %
is $912\times$
(Table~\ref{tab:rtf}).
With WaveGlow, RTF for complete audio synthesis drops down to $63\times$.
RTF measured on Intel Xeon Gold 6240 CPU is $108\times$.
FastPitch is suitable to real-time editing of synthesized samples,
and a single pitch value per input symbol is easy to interpret by a human,
which we demonstrate in a video clip on the aforementioned website with samples.
\begin{table}[htb]
    \centering
    \caption{Average latency and real-time factor of mel-spectrogram generation on the first 2048 utterances from LJSpeech training subset}  %
    \label{tab:rtf}
    \begin{tabular}{l l r}
      \toprule
        Model                         & Latency & Mel RTF\\
      \midrule
        Tacotron 2 (GPU) & $0.4109 \pm \SI{0.2631}{\second}$ & $15.42\times$\\
        FastPitch (CPU) & $0.0602 \pm \SI{0.0205}{\second}$ & $107.59\times$\\
        FastPitch (GPU) & $0.0071\pm \SI{0.0010}{\second}$ & $911.86\times$\\
      \bottomrule
    \end{tabular}
\end{table}

\section{CONCLUSIONS}
We have presented FastPitch, a parallel text-to-speech model based on FastSpeech,
able to rapidly synthesize high-fidelity mel-scale spectrograms with a high
degree of control over the prosody.
The model demonstrates how conditioning on prosodic
information can significantly improve the convergence and quality of synthesized speech
in a feed-forward model, enabling more coherent pronunciation
across its independent outputs, and lead to state-of-the-art results.
Our pitch conditioning method is simpler
than many of the approaches known from the literature.
It does not introduce an overhead,
and opens up possibilities for practical applications in adjusting the prosody interactively,
as the model is fast, highly expressive, and presents potential for multi-speaker scenarios.

\section{ACKNOWLEDGEMENTS}
The author would like to thank
Dabi Ahn,
Alvaro Garcia,
and
Grzegorz Karch
for their help with the experiments and evaluation of the model,
and
Jan Chorowski,
João Felipe Santos,
Przemek Strzelczyk,
and
Rafael Valle
for helpful discussions and support in
preparation of this paper.

\vfill\pagebreak

\bibliographystyle{IEEEbib}
\bibliography{refs}

\end{document}